\documentclass[a4paper,11pt]{article}
\usepackage{pos}

\title{Heavy-flavor production and hadronization at the LHC: experimental status and perspectives from LHC experiments}
\ShortTitle{Heavy-flavor production and hadronization at the LHC}

\author*[a]{Victor Feuillard}

\onbehalf{on behalf of the ALICE, ATLAS, CMS and LHCb collaborations}

\affiliation[a]{Heidelberg University,\\
  Im Neuenheimer Feld 226, Heidelberg, Germany}

\emailAdd{victor.jose.gaston.feuillard@cern.ch}

\abstract{Heavy-flavor hadrons are one of the most prominent probes to study the quark--gluon plasma and to test models based on Quantum Chromodynamics (QCD). This contribution presents the latest results regarding heavy-flavor production in ALICE, ATLAS, CMS and LHCb.}

\FullConference{%
12th Large Hadron Collider Physics Conference -- LHCP 2024 \\
3-8 June 2024 \\
Boston, MA, US
}

\begin{document}
\maketitle

Heavy-flavor hadrons contain at least one charm or beauty valence quark. Because of the large masses of the heavy quarks, their formation time is short and, in heavy-ion collisions, they experience the whole evolution of the quark--gluon plasma (QGP) medium produced in the collision. Heavy quarks are produced in initial hard scatterings with moderate to large $Q^2$ and their production can be described with perturbative quantum chromodynamics (pQCD) calculations, using the factorization approach, in which the production cross-section is proportional to the parton distribution functions, the partonic cross section, and the fragmentation functions. The latter were assumed for a long time to be universal across collision systems but new results indicate otherwise. 

Heavy-flavor production can be studied in several collision systems, with different goals. Measurements in proton--proton collisions allow us to test pQCD calculations describing heavy-flavor hadron production and measure the heavy-quark fragmentation functions. Measurements in proton--nucleus collisions are helpful to investigate initial-state effects such as shadowing and gluon saturation, as well as study the interplay between soft and hard processes. Finally, measurements in nucleus--nucleus collisions probe the properties of the QGP and measure the influence of final-state effects in heavy-flavor production. In the following we will present some of the latest results measured by ALICE, ATLAS, CMS and LHCb on heavy-flavor production and hadronization.

\begin{figure}[!htb]
\begin{center}
\vspace{9pt}
\includegraphics[scale=0.37]{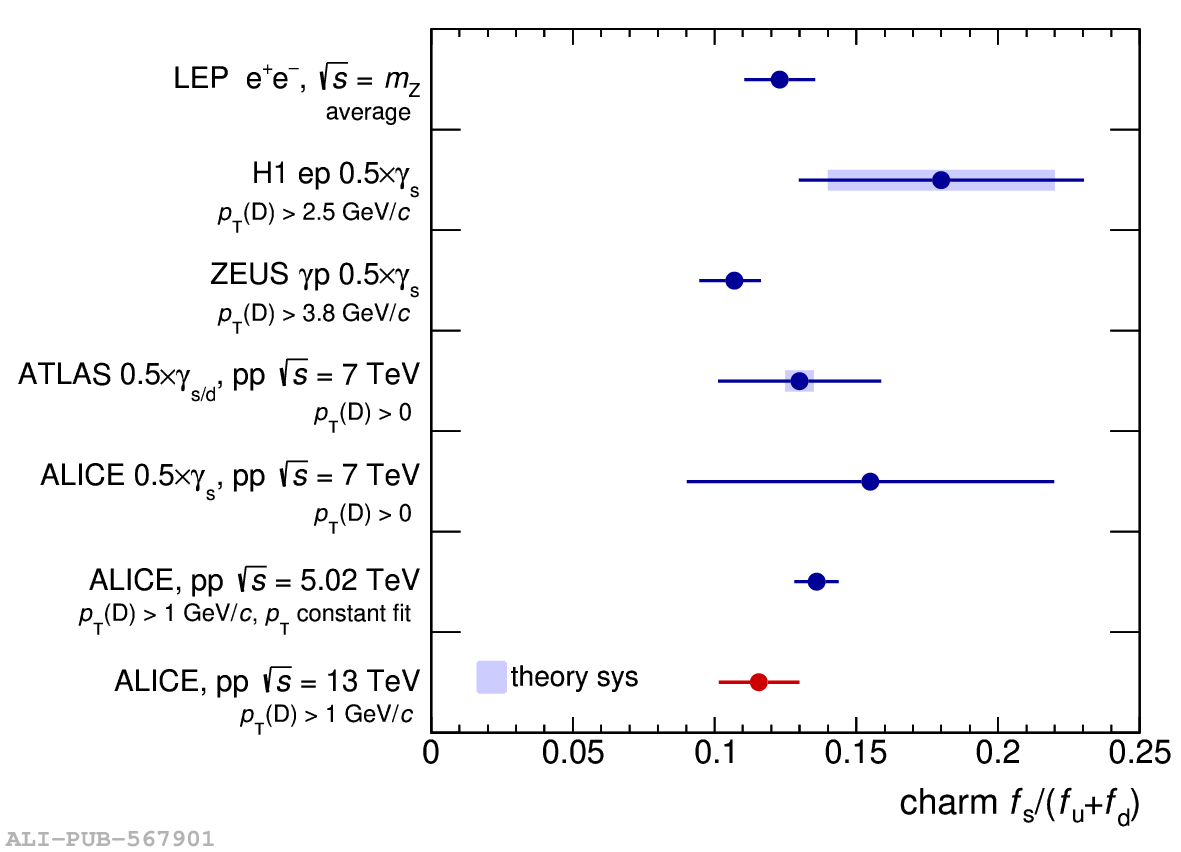}
\includegraphics[scale=0.37]{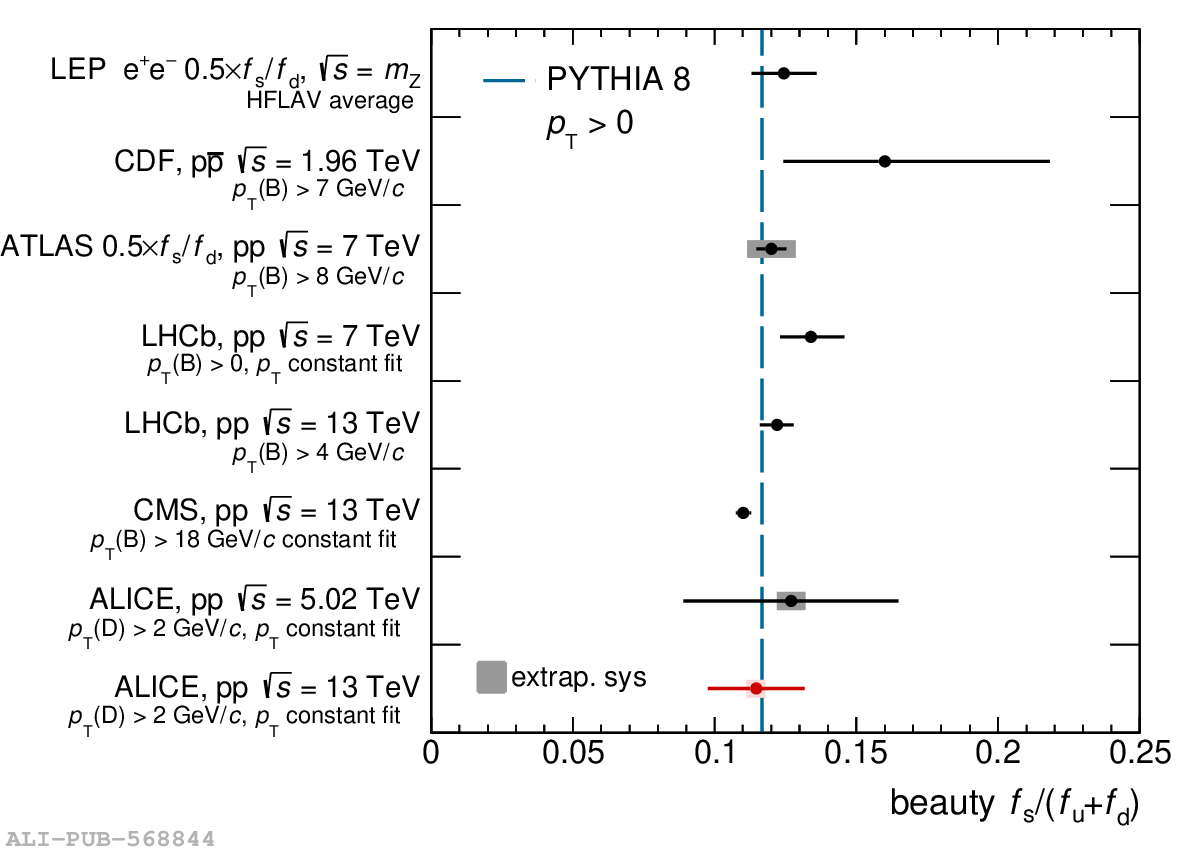}
\caption{Left: Ratio of strange-to-non-strange charm fragmentation functions measured by ALICE in pp collisions at $\sqrt{s}=13$~TeV compared with other experiments~\cite{DMesonPromptALICE}. Right: Ratio of strange-to-non-strange beauty fragmentation functions measured by ALICE in pp collisions at $\sqrt{s}=13$~TeV compared with other experiments~\cite{DMesonNonPromptALICE}.}
\label{fig:Figure1}
\end{center}
\end{figure}

The production of prompt and non-prompt D mesons has been measured by ALICE in pp collisions at $\sqrt{s}=13$~TeV~\cite{DMesonPromptALICE,DMesonNonPromptALICE}. The strange-to-non-strange production yield ratio, D$_{\rm s}^+$/(D$^0$ + D$^+$), has been measured as a function of $p_{\rm T}$. In the prompt case, the ratio exhibits an increasing trend as a function of $p_{\rm T}$ up to around 8~GeV/$c$, while no significant trend is visible in the non-prompt case due to the larger uncertainties. The non-prompt measurement has been compared with FONLL pQCD calculations~\cite{DMesonNonPromptFONLL}, which are able to describe the data in the whole $p_{\rm T}$ range. From these measurements the strange-to-non-strange ratio of fragmentation functions can be extracted. The corresponding result is presented in Figure~\ref{fig:Figure1} for the charm (left) and beauty (right) cases, and compared with measurements by other experiments and in other collision systems. All measurements are in agreement within uncertainties, indicating a universality of the relative fragmentation function for charm and beauty mesons.

The strange-to-non-strange production yield ratio for beauty has also been measured in Pb--Pb collisions for the B mesons by CMS~\cite{BmesonCMSPbPb} and non-prompt D mesons by ALICE~\cite{DmesonALICEPbPb,DmesonALICEPbPb2,DmesonALICEPbPb3}. These results show a hint that strange mesons are less suppressed than non-strange mesons, as is expected in the presence of strangeness enhancement inside a QGP. A transport model implementing strangeness enhancement and hadronization via recombination~\cite{TAMUStrangeRatio} is compatible with data, as well as predictions from a statistical hadronization model~\cite{SHMStrangeRatio}. However, due to the large uncertainties on the data, the measurements are also compatible with a scenario without strangeness enhancement.

\begin{figure}[!htb]
\begin{center}
\vspace{9pt}
\includegraphics[scale=0.33]{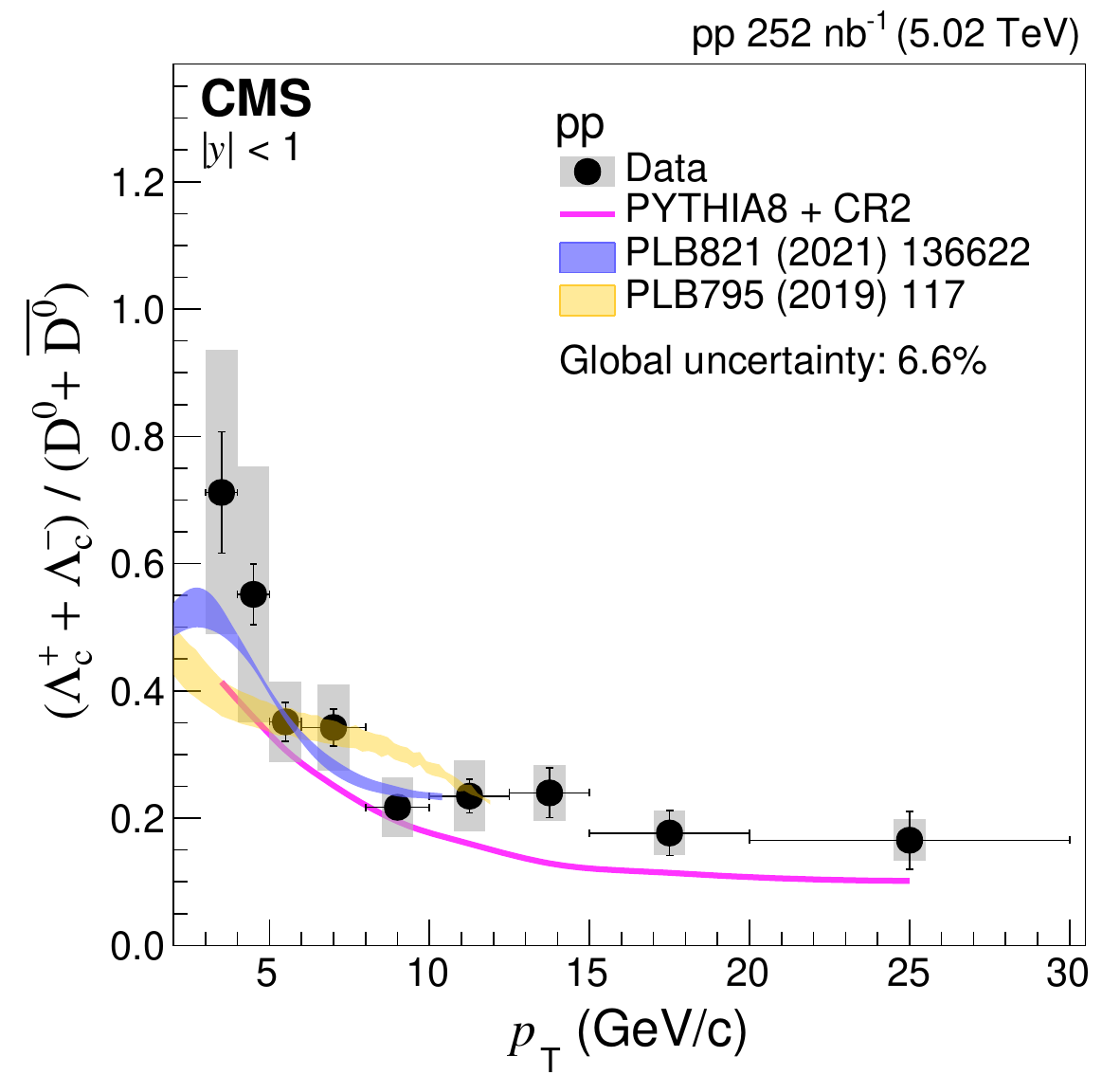}
\includegraphics[scale=0.42]{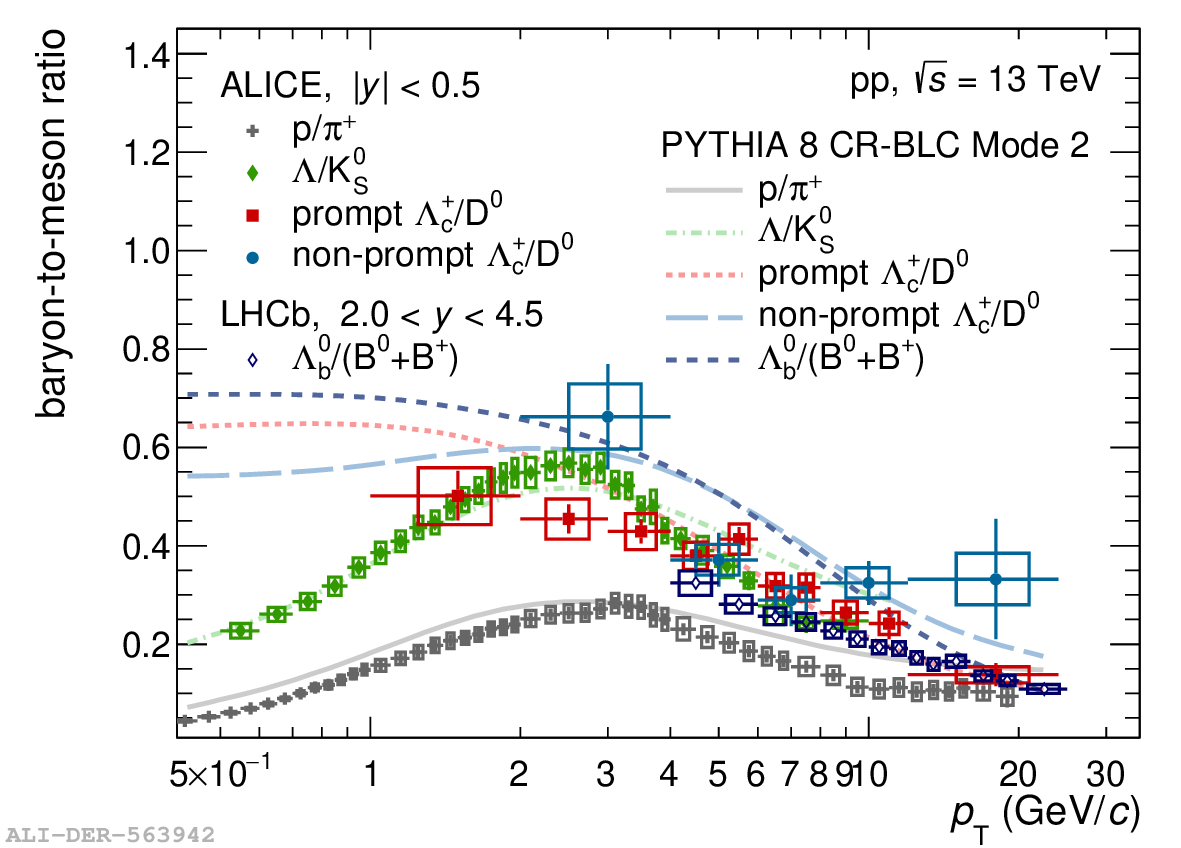}
\caption{Left: Prompt and non-prompt $\Lambda_{\rm{c}}^{+}$ over D$^0$ production yield ratio measured by CMS in pp collisions at $\sqrt{s}=5.02$~TeV~\cite{LambdaCCMS} compared with different models~\cite{Pythia8LamdbaC,CataniaLambdaC,SHMLambdaC}. Right: $\Lambda_{\rm{c}}^{+}$ over D$^0$ production ratio measured by ALICE in pp collisions at $\sqrt{s}=13$~TeV~\cite{LambdaCALICE} compared with PYTHIA calculations~\cite{Pythia8LamdbaC} and LHCb measurements of the $\Lambda_{\rm{b}}^{+}$/B yield ratio~\cite{LambdaCLHCb}.}
\label{fig:Figure2}
\end{center}
\end{figure}

The $\Lambda_{\rm{c}}^{+}$-baryon production has been measured at mid-rapidity by CMS in pp collisions at $\sqrt{s}=5.02$~TeV~\cite{LambdaCCMS} and ALICE in pp collisions at $\sqrt{s}=13$~TeV~\cite{LambdaCALICE} for $p_{\rm T}>3$~GeV/$c$ and $p_{\rm T}>1$~GeV/$c$, respectively. The $\Lambda_{\rm{c}}^{+}$ over D$^0$ yield ratio is presented in Figure~\ref{fig:Figure2}, compared with earlier measurements of the $\Lambda_{\rm{b}}^{+}$ over B ratio from LHCb~\cite{LambdaCLHCb} at forward rapidity.

In both CMS and ALICE, the $\Lambda_{\rm{c}}^{+}$ over D$^0$ production ratio shows a decreasing trend with increasing $p_{\rm T}$, differently from what was measured in leptonic collisions, where no significant $p_{\rm T}$ dependence was observed. PYTHIA8~\cite{Pythia8LamdbaC} calculations implementing a specific tune with color reconnection beyond leading color (BLC-CR mode 2) show a good agreement with the data. The non-prompt $\Lambda_{\rm{c}}^{+}$ over D$^0$ ratio measured by ALICE is generally higher than the $\Lambda_{\rm{b}}^{+}$ over B ratio measured by LHCb. Models with coalescence and fragmentation processes~\cite{CataniaLambdaC} show a good agreement in the available $p_{\rm T}$ range and reproduce the decreasing trend over $p_{\rm T}$. The statistical hadronization model~\cite{SHMLambdaC} also shows a good agreement with the data in the available $p_{\rm T}$ range. 

\begin{figure}[!htb]
\begin{center}
\vspace{9pt}
\includegraphics[scale=0.40]{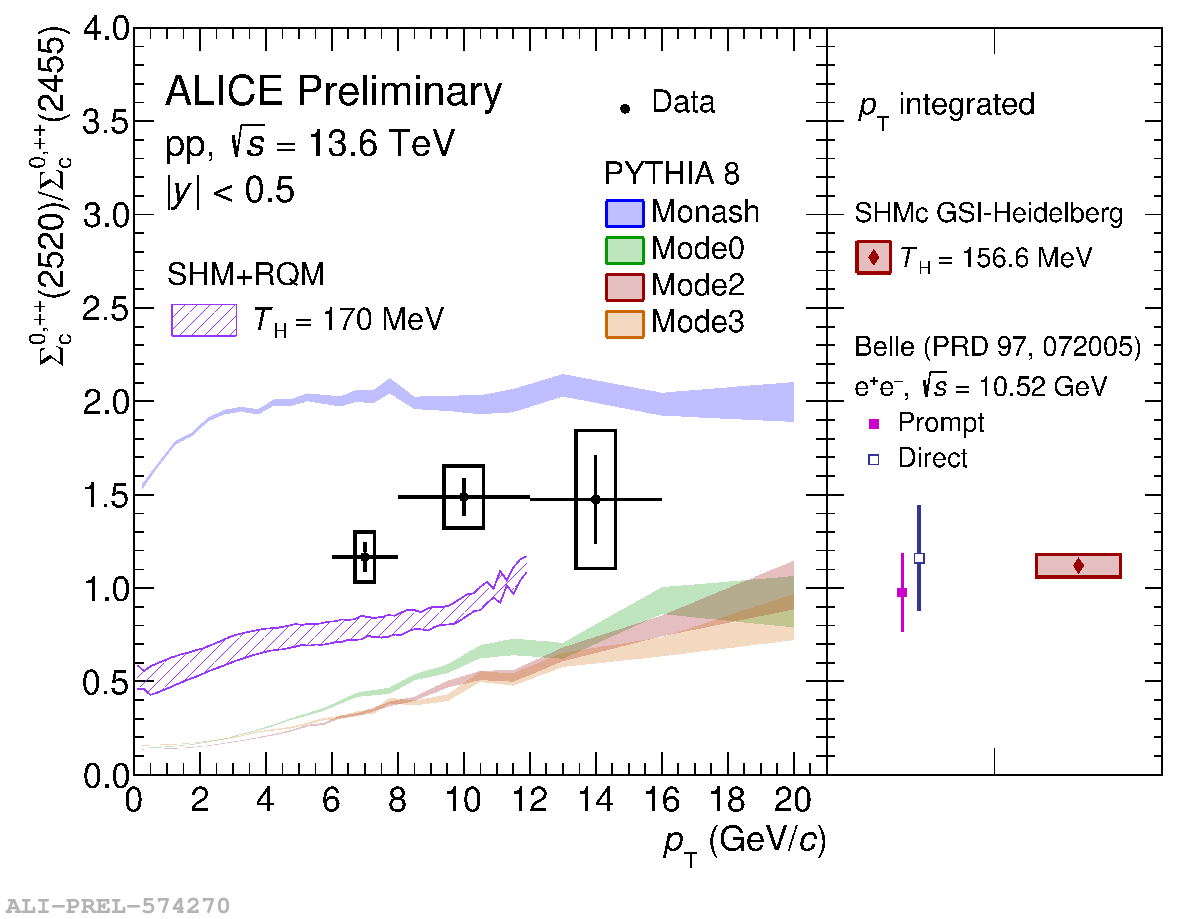}
\caption{Measurement of the $\Sigma_{\rm c}^{0,++}(2520)$/$\Sigma_{\rm c}^{0,++}(2455)$ yield ratio in ALICE compared with models~\cite{SigmaCPythia,SigmaCSHMRQM,SigmaCSHMc} and Belle measurement~\cite{SigmaCBelle}. }
\label{fig:Figure21}
\end{center}
\end{figure}

These measurements are extended to additional states by new results coming from Run 3 of the LHC, as illustrated by the first measurement of the $\Sigma_{\rm c}^{0,++}(2520)$ in ALICE presented in Figure~\ref{fig:Figure21}. The $\Sigma_{\rm c}^{0,++}(2520)$/$\Sigma_{\rm c}^{0,++}(2455)$ ratio is consistent with the $p_{\rm T}$-integrated result from e$^+$e$^-$ collisions within the uncertainties in the investigated $p_{\rm T}$ region. The result is compared with several calculations: PYTHIA 8~\cite{SigmaCPythia} Monash overestimates the data, while the CR Modes underestimate the data. The Statistical Hadronization Model~\cite{SigmaCSHMRQM} prediction also slightly underestimates the data in the common $p_{\rm T}$ range.

\begin{figure}[!htb]
\begin{center}
\vspace{9pt}
\includegraphics[scale=0.35]{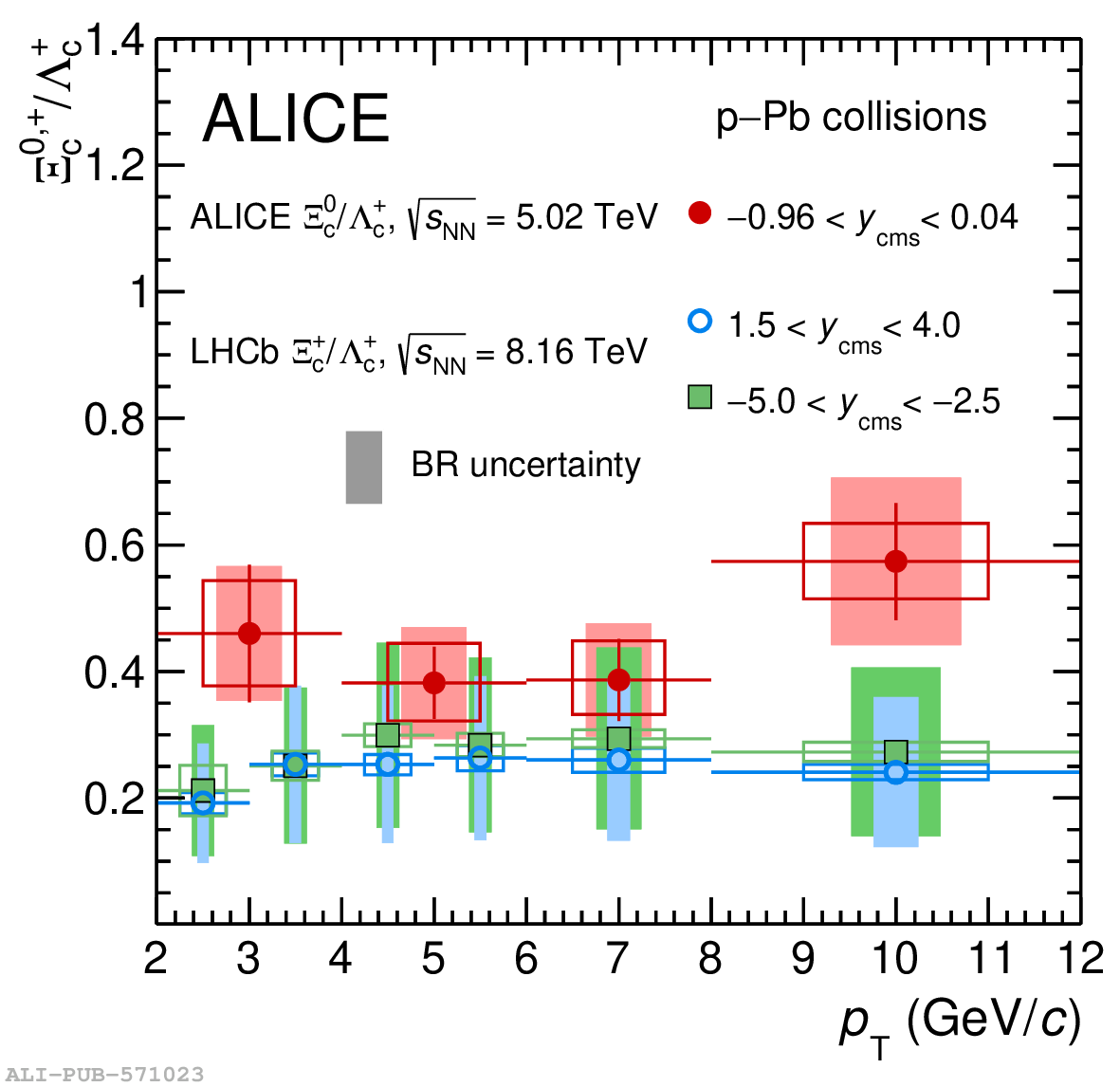}
\includegraphics[scale=0.40]{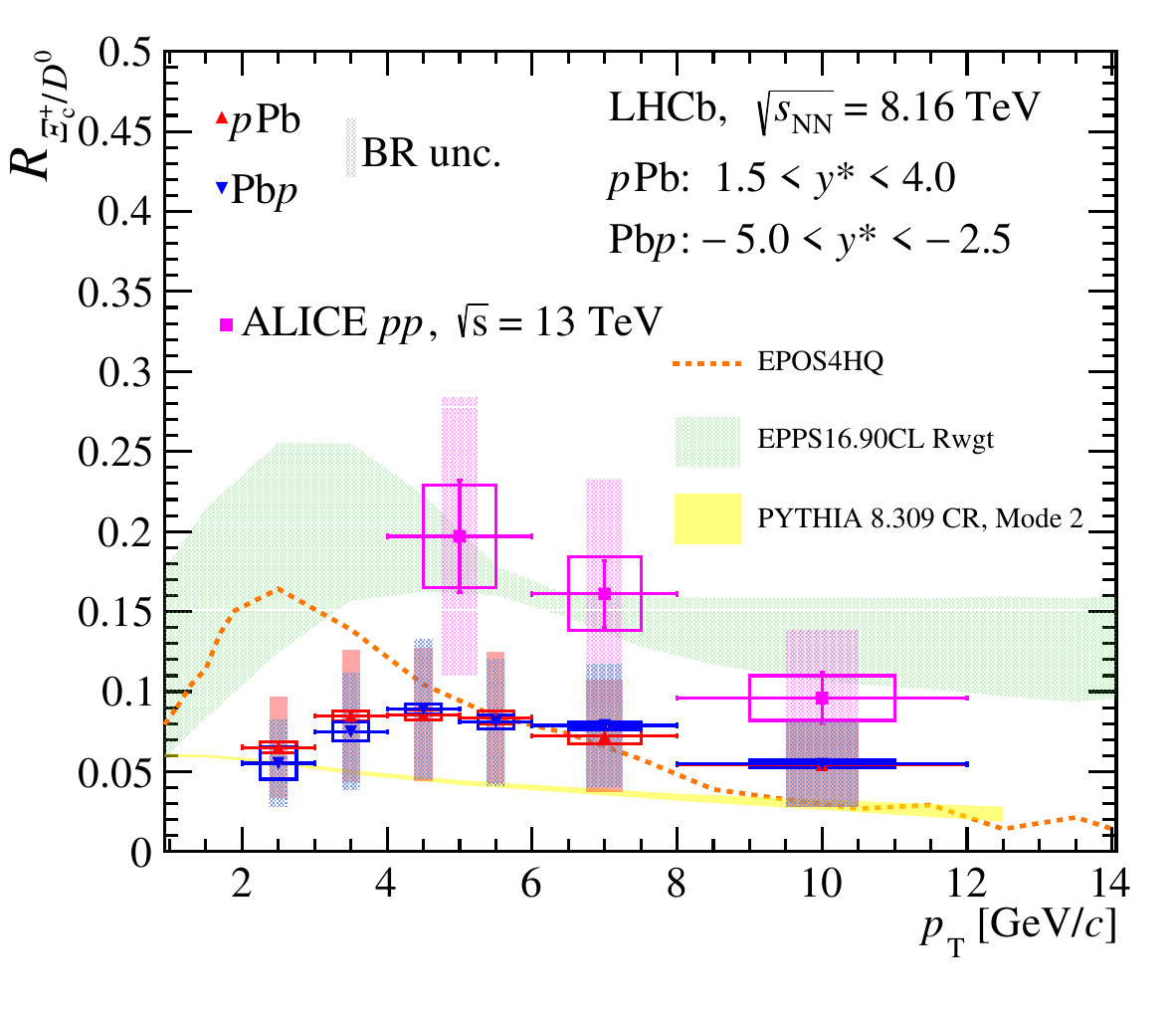}
\caption{Left:  $\Xi_{\rm{c}}^{+}$/$\Lambda_{\rm{c}}^{+}$  ratio measured by ALICE at $\sqrt{s_{\rm{NN}}}=5.02$~TeV~\cite{ChiCALICE} and  LHCb at $\sqrt{s_{\rm{NN}}}=8.16$~TeV~\cite{ChiCLHCb} Right: $\Xi_{\rm{c}}^{+}$/D$^0$ ratio measured LHCb at $\sqrt{s_{\rm{NN}}}=8.16$~TeV~\cite{ChiCLHCb} compared with ALICE measurements in pp collisions~\cite{ChiCALICEpp} and EPSS16~\cite{EPPS16DmesonpPb}, PYTHIA~\cite{Pythia8LamdbaC} and EPOS~\cite{EPOS} predictions.}
\label{fig:Figure3}
\end{center}
\end{figure}

The $\Xi_{\rm{c}}^{+}$/D$^0$ and $\Xi_{\rm{c}}^{+}$/$\Lambda_{\rm{c}}^{+}$  production yield ratios have been measured in p--Pb collisions by ALICE at $\sqrt{s_{\rm{NN}}}=5.02$~TeV~\cite{ChiCALICE} and by LHCb at $\sqrt{s_{\rm{NN}}}=8.16$~TeV~\cite{ChiCLHCb}. The results are presented in Figure~\ref{fig:Figure3}.

The results from ALICE and LHCb are compatible within the uncertainties, although the ALICE result tend to be higher. The ratios show no significant $p_{\rm T}$ dependence for both p-going and Pb-going directions in both ALICE and LHCb, which is a strong indication that the same processes govern hadronization in forward and backward rapidity. The LHCb data are also compared with models (Figure~\ref{fig:Figure3} right). The EPPS16~\cite{EPPS16DmesonpPb} predictions overestimate the LHCb data but show a similar trend, while agreeing with the ALICE measurements. PYTHIA8.3 BLC-CR mode 2~\cite{Pythia8LamdbaC} calculations lie on the lower edge of the data uncertainties, and EPOS4HQ~\cite{EPOS} calculations describe the data but show a different trend as a function of $p_{\rm T}$ with respect to the LHCb measurement.

\begin{figure}[!htb]
\begin{center}
\vspace{9pt}
\includegraphics[scale=0.45]{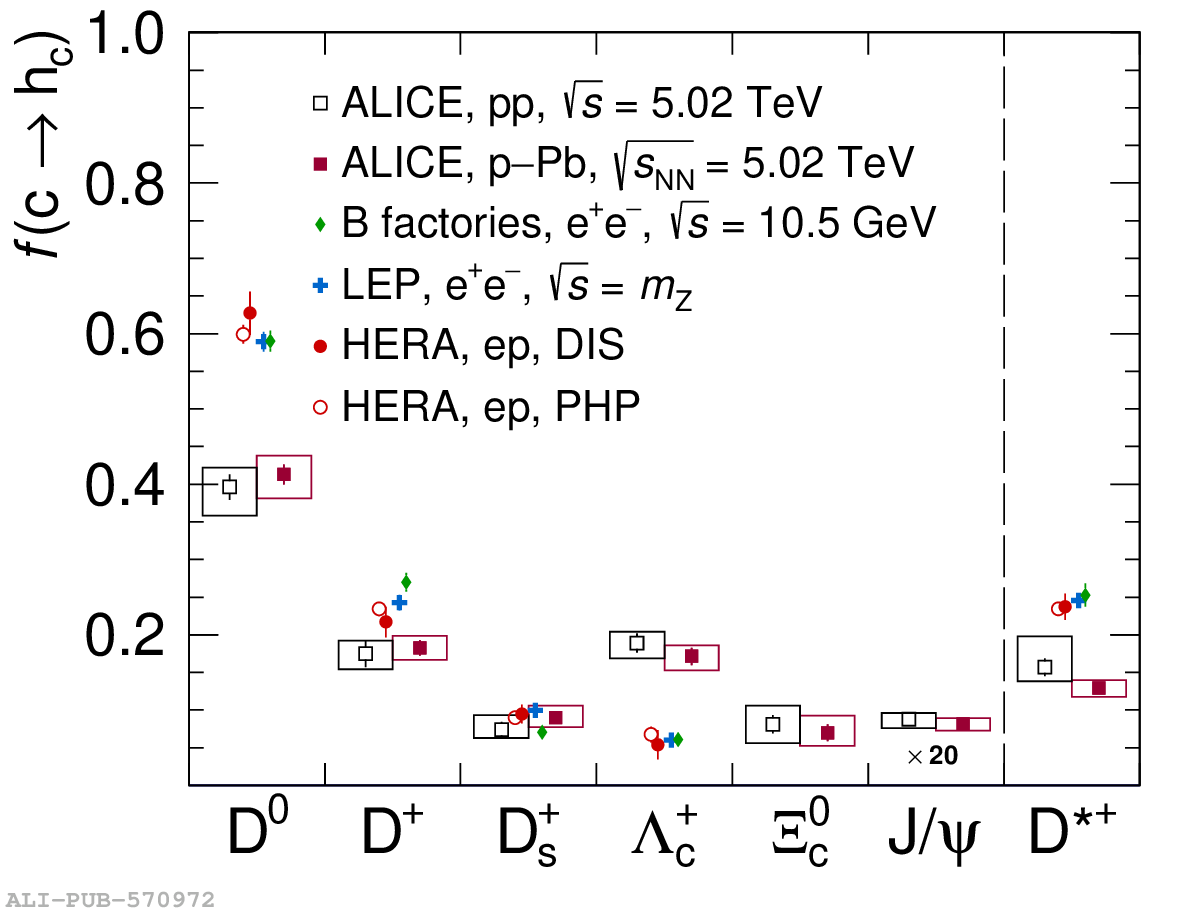}
\caption{Charm fragmentation fraction measured by ALICE in pp and p--Pb collisions compared with other collision systems~\cite{FragFunctionsALICE}.}
\label{fig:Figure4}
\end{center}
\end{figure}

All the measurements of heavy-flavor mesons and baryons were used to evaluate the charm fragmentation fractions~\cite{FragFunctionsALICE}, which are shown in Figure~\ref{fig:Figure4}. The values obtained by ALICE are consistent between pp and p--Pb collisions. However a large difference is observed with respect to e$^+$e$^-$ and ep collisions: an increase in the $\Lambda_{\rm{c}}^{+}$  production is accompanied by a concomitant decrease in non-strange D-meson production. This indicates that the universality of parton-to-hadron fragmentation is not generally valid, contrary to what was assumed until now.

\begin{figure}[!htb]
\begin{center}
\vspace{9pt}
\includegraphics[scale=0.90]{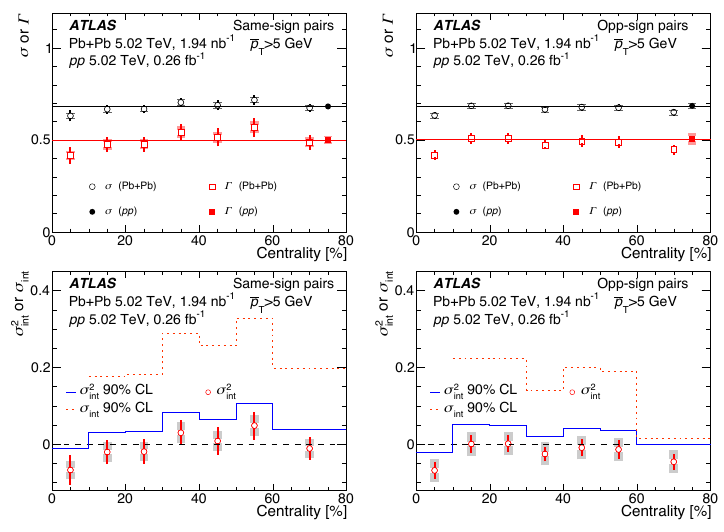}
\caption{Two-muon correlation function peak widths as a function of centrality, in pp and Pb--Pb collisions at  $\sqrt{s_{\rm{NN}}}=5.02$~TeV~\cite{MuonsATLAS}.}
\label{fig:Figure5}
\end{center}
\end{figure}

The ATLAS experiment has measured the two-muon correlation functions in pp collisions at $\sqrt{s} = 5.02$~TeV and Pb--Pb collisions at $\sqrt{s_{\rm{NN}}}=5.02$~TeV as a function of the azimuthal angle difference $\Delta \phi$~\cite{MuonsATLAS}. As expected for semileptonic decays of heavy-quark pairs, a strong peak is visible at $\Delta \phi = \pi$, with a width that can be quantified by a FWHM $\Gamma$ or a standard deviation $\sigma$, on top of an uncorrelated pedestal. The peak widths  are shown in Figure~\ref{fig:Figure5}, and are the same for the same-sign and opposite-sign pairs, indicating a dominance of beauty decays in this kinematic region. The widths in Pb--Pb collisions are the same as in pp collisions, except for the most central collisions, which show a narrower peak. This could be an effect of a different $p_{\rm T}$ distribution of the originating b quarks, which is modified by the energy loss of the quarks in the QGP, but does not support the predicted systematic increase in the $\Gamma$ broadening from peripheral to central collisions due to interactions in the QGP~\cite{MuodelATLAS}. The bottom panel of Figure~\ref{fig:Figure5} shows that the square of such an additional broadening in Pb--Pb collisions with respect to pp collisions is negligible and even results in a negative contribution at 90\% confidence level in the most central interval.

\begin{figure}[!htb]
\begin{center}
\vspace{9pt}
\includegraphics[scale=0.39]{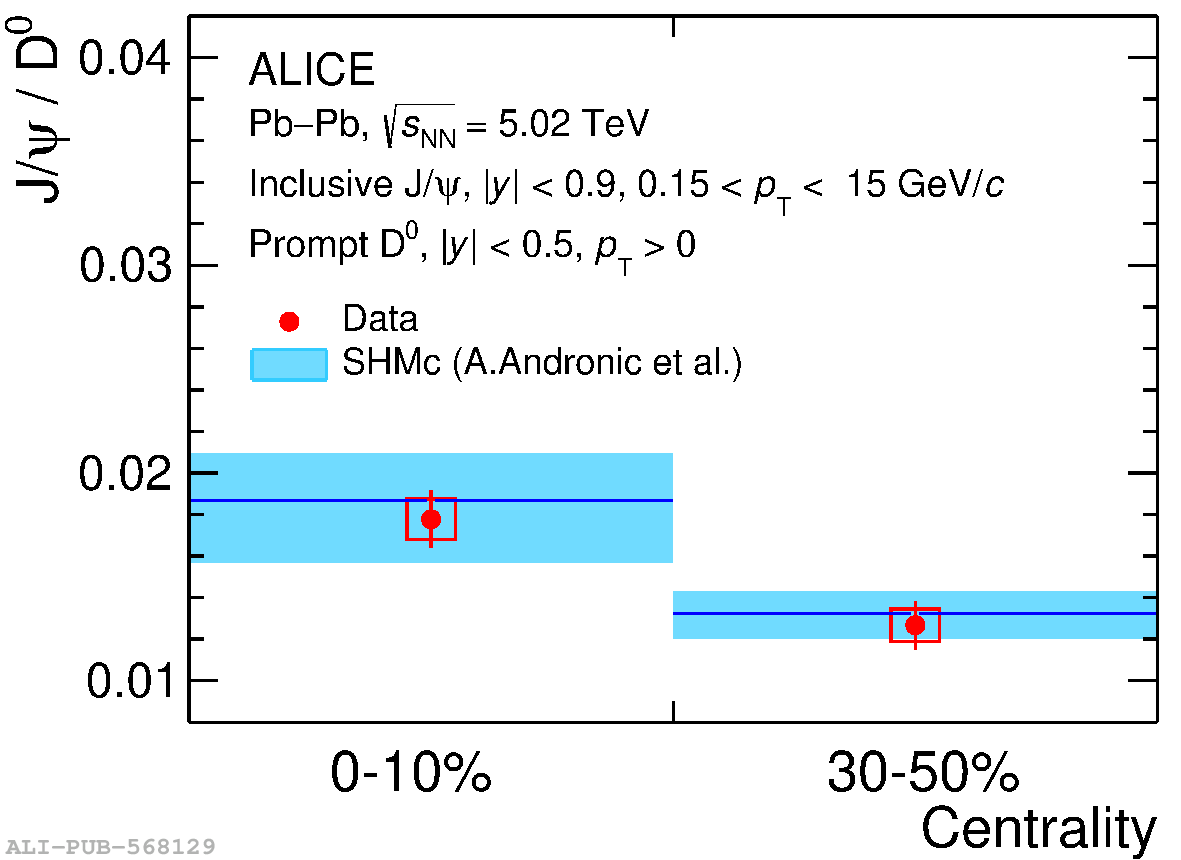}
\caption{J/$\psi$/D$^0$ ratio as a function of centrality in Pb--Pb collisions at $\sqrt{s_{\rm{NN}}}=5.02$~TeV~\cite{JpsiDALICE} compared with SHMc predictions~\cite{SigmaCSHMc}.}
\label{fig:Figure6}
\end{center}
\end{figure}

Finally, the J/$\psi$/D$^0$ ratio has been measured by ALICE in Pb--Pb collisions at $\sqrt{s_{\rm{NN}}}=5.02$~TeV~\cite{JpsiDALICE}. This ratio provides a tight constraint to models because it simultaneously describes open and hidden charm production using only the measured charm production cross-section, and is particularly sensitive to the hadronization mechanisms of the different charm hadrons. The result is presented in Figure~\ref{fig:Figure6} as a function of centrality compared with SHMc model predictions~\cite{SigmaCSHMc}. The ratio is higher in most central collision interval, a feature correctly described by the SHMc model. This hints that both J/$\psi$ and D$^0$ are produced via the statistical hadronization of deconfined and locally thermalized charm quarks.

In conclusion, the presented results show that the measurement of multiple heavy-flavor species offers a solid ground to test pQCD models and the factorization approach, and suggests a breaking of universal hadronization across different collision systems. The ability to measure several observables ($R_{\rm{AA }}$, production ratios...) provides many avenues for model comparison and improves our understanding of heavy-quark interaction with the medium. Measurements indicate that, in Pb--Pb collisions, charm hadrons can also produced through coalescence or statistical recombination at low and moderate $p_{\rm T}$, whereas no significant contribution from coalescence is observed at high  $p_{\rm T}$ in Pb--Pb collisions. With new results coming out of the ongoing Run 3 of the LHC, more precise measurements with reduced uncertainties will be able to further improve our understanding of heavy-flavor production and hadronization.

\begingroup
    \setlength{\bibsep}{1pt}
    \linespread{1}\selectfont
    \begin{scriptsize}
\bibliographystyle{JHEP.bst}
\bibliography{bibliography.bib}

\providecommand{\href}[2]{#2}\begingroup\raggedright\begin{thebibliography}{10}

\bibitem{DMesonPromptALICE}
{ ALICE} Collaboration,
  \href{https://doi.org/10.1007/JHEP12(2023)086}{\emph{JHEP} {\bfseries 12}
  (2023) 086} [\href{https://arxiv.org/abs/2308.04877}{{\ttfamily
  2308.04877}}].

\bibitem{DMesonNonPromptALICE}
{ ALICE} Collaboration,
  \href{https://doi.org/10.1007/JHEP10(2024)110}{\emph{JHEP} {\bfseries 10}
  (2024) 110} [\href{https://arxiv.org/abs/2402.16417}{{\ttfamily
  2402.16417}}].

\bibitem{DMesonNonPromptFONLL}
M.~Cacciari et~al., \href{https://doi.org/10.1007/JHEP10(2012)137}{\emph{JHEP}
  {\bfseries 10} (2012) 137} [\href{https://arxiv.org/abs/1205.6344}{{\ttfamily
  1205.6344}}].

\bibitem{BmesonCMSPbPb}
{ CMS} Collaboration,
  \href{https://doi.org/10.1016/j.physletb.2022.137062}{\emph{Phys. Lett. B}
  {\bfseries 829} (2022) 137062}
  [\href{https://arxiv.org/abs/2109.01908}{{\ttfamily 2109.01908}}].

\bibitem{DmesonALICEPbPb}
{ ALICE} Collaboration,
  \href{https://doi.org/10.1016/j.physletb.2022.137561}{\emph{Phys. Lett. B}
  {\bfseries 846} (2023) 137561}
  [\href{https://arxiv.org/abs/2204.10386}{{\ttfamily 2204.10386}}].

\bibitem{DmesonALICEPbPb2}
{ ALICE} Collaboration,
  \href{https://doi.org/10.1016/j.physletb.2022.136986}{\emph{Phys. Lett. B}
  {\bfseries 827} (2022) 136986}
  [\href{https://arxiv.org/abs/2110.10006}{{\ttfamily 2110.10006}}].

\bibitem{DmesonALICEPbPb3}
{ ALICE} Collaboration,
  \href{https://doi.org/10.1007/JHEP01(2022)174}{\emph{JHEP} {\bfseries 01}
  (2022) 174} [\href{https://arxiv.org/abs/2110.09420}{{\ttfamily
  2110.09420}}].

\bibitem{TAMUStrangeRatio}
M.~He, R.J.~Fries and R.~Rapp,
  \href{https://doi.org/10.1016/j.physletb.2014.05.050}{\emph{Phys. Lett. B}
  {\bfseries 735} (2014) 445}
  [\href{https://arxiv.org/abs/1401.3817}{{\ttfamily 1401.3817}}].

\bibitem{SHMStrangeRatio}
P.~Braun-Munzinger, K.~Redlich, N.~Sharma and J.~Stachel,
  \href{https://arxiv.org/abs/2408.07496}{{\ttfamily 2408.07496}}.

\bibitem{LambdaCCMS}
{ CMS} Collaboration,
  \href{https://doi.org/10.1007/JHEP01(2024)128}{\emph{JHEP} {\bfseries 01}
  (2024) 128} [\href{https://arxiv.org/abs/2307.11186}{{\ttfamily
  2307.11186}}].

\bibitem{LambdaCALICE}
{ ALICE} Collaboration,
  \href{https://doi.org/10.1103/PhysRevD.108.112003}{\emph{Phys. Rev. D}
  {\bfseries 108} (2023) 112003}
  [\href{https://arxiv.org/abs/2308.04873}{{\ttfamily 2308.04873}}].

\bibitem{LambdaCLHCb}
{ LHCb} Collaboration,
  \href{https://doi.org/10.1103/PhysRevD.100.031102}{\emph{Phys. Rev. D}
  {\bfseries 100} (2019) 031102}
  [\href{https://arxiv.org/abs/1902.06794}{{\ttfamily 1902.06794}}].

\bibitem{Pythia8LamdbaC}
J.R.~Christiansen and P.Z.~Skands,
  \href{https://doi.org/10.1007/JHEP08(2015)003}{\emph{JHEP} {\bfseries 08}
  (2015) 003} [\href{https://arxiv.org/abs/1505.01681}{{\ttfamily
  1505.01681}}].

\bibitem{CataniaLambdaC}
V.~Minissale, S.~Plumari and V.~Greco,
  \href{https://doi.org/10.1016/j.physletb.2021.136622}{\emph{Phys. Lett. B}
  {\bfseries 821} (2021) 136622}
  [\href{https://arxiv.org/abs/2012.12001}{{\ttfamily 2012.12001}}].

\bibitem{SHMLambdaC}
M.~He and R.~Rapp,
  \href{https://doi.org/10.1016/j.physletb.2019.06.004}{\emph{Phys. Lett. B}
  {\bfseries 795} (2019) 117}
  [\href{https://arxiv.org/abs/1902.08889}{{\ttfamily 1902.08889}}].

\bibitem{SigmaCPythia}
J.R.~Christiansen and P.Z.~Skands,
  \href{https://doi.org/10.1007/JHEP08(2015)003}{\emph{JHEP} {\bfseries 08}
  (2015) 003} [\href{https://arxiv.org/abs/1505.01681}{{\ttfamily
  1505.01681}}].

\bibitem{SigmaCSHMRQM}
M.~He and R.~Rapp,
  \href{https://doi.org/10.1016/j.physletb.2019.06.004}{\emph{Phys. Lett. B}
  {\bfseries 795} (2019) 117}
  [\href{https://arxiv.org/abs/1902.08889}{{\ttfamily 1902.08889}}].

\bibitem{SigmaCSHMc}
A.~Andronic, P.~Braun-Munzinger, M.K.~K\"ohler, K.~Redlich and J.~Stachel,
  \href{https://doi.org/10.1016/j.physletb.2019.134836}{\emph{Phys. Lett. B}
  {\bfseries 797} (2019) 134836}
  [\href{https://arxiv.org/abs/1901.09200}{{\ttfamily 1901.09200}}].

\bibitem{SigmaCBelle}
{ Belle} Collaboration,
  \href{https://doi.org/10.1103/PhysRevD.97.072005}{\emph{Phys. Rev. D}
  {\bfseries 97} (2018) 072005}
  [\href{https://arxiv.org/abs/1706.06791}{{\ttfamily 1706.06791}}].

\bibitem{ChiCALICE}
{ ALICE} Collaboration,  \href{https://arxiv.org/abs/2405.14538}{{\ttfamily
  2405.14538}}.

\bibitem{ChiCLHCb}
{ LHCb} Collaboration,
  \href{https://doi.org/10.1103/PhysRevC.109.044901}{\emph{Phys. Rev. C}
  {\bfseries 109} (2024) 044901}
  [\href{https://arxiv.org/abs/2305.06711}{{\ttfamily 2305.06711}}].

\bibitem{ChiCALICEpp}
{ ALICE} Collaboration,
  \href{https://doi.org/10.1103/PhysRevLett.127.272001}{\emph{Phys. Rev. Lett.}
  {\bfseries 127} (2021) 272001}
  [\href{https://arxiv.org/abs/2105.05187}{{\ttfamily 2105.05187}}].

\bibitem{EPPS16DmesonpPb}
K.J.~Eskola, P.~Paakkinen, H.~Paukkunen and C.A.~Salgado,
  \href{https://doi.org/10.1140/epjc/s10052-017-4725-9}{\emph{Eur. Phys. J. C}
  {\bfseries 77} (2017) 163}
  [\href{https://arxiv.org/abs/1612.05741}{{\ttfamily 1612.05741}}].

\bibitem{EPOS}
K.~Werner and B.~Guiot,
  \href{https://doi.org/10.1103/PhysRevC.108.034904}{\emph{Phys. Rev. C}
  {\bfseries 108} (2023) 034904}
  [\href{https://arxiv.org/abs/2306.02396}{{\ttfamily 2306.02396}}].

\bibitem{FragFunctionsALICE}
{ ALICE} Collaboration,  \href{https://arxiv.org/abs/2405.14571}{{\ttfamily
  2405.14571}}.

\bibitem{MuonsATLAS}
{ ATLAS} Collaboration,
  \href{https://doi.org/10.1103/PhysRevLett.132.202301}{\emph{Phys. Rev. Lett.}
  {\bfseries 132} (2024) 202301}
  [\href{https://arxiv.org/abs/2308.16652}{{\ttfamily 2308.16652}}].

\bibitem{MuodelATLAS}
M.~Nahrgang, J.~Aichelin, P.B.~Gossiaux and K.~Werner,
  \href{https://doi.org/10.1103/PhysRevC.90.024907}{\emph{Phys. Rev. C}
  {\bfseries 90} (2014) 024907}
  [\href{https://arxiv.org/abs/1305.3823}{{\ttfamily 1305.3823}}].

\bibitem{JpsiDALICE}
{ ALICE} Collaboration,
  \href{https://doi.org/10.1016/j.physletb.2024.138451}{\emph{Phys. Lett. B}
  {\bfseries 849} (2024) 138451}
  [\href{https://arxiv.org/abs/2303.13361}{{\ttfamily 2303.13361}}].

\end{thebibliography}\endgroup
\end{scriptsize}
\endgroup

\end{document}